\documentclass[preprint]{aastex}

\slugcomment{To appear in ApJ (scheduled for the v555 n2 Jul 10, 2001 issue)}

\shorttitle{Momose et al. }
\shortauthors{Imaging Polarimetry of NGC 7538}

\begin{document}

\title{Submillimeter Imaging Polarimetry of the NGC 7538 Region}

\author{M. Momose\altaffilmark{1,2}, M. Tamura\altaffilmark{3}, 
O. Kameya\altaffilmark{3,4}, J. S. Greaves\altaffilmark{5}, 
A. Chrysostomou\altaffilmark{5,6}, \\
J. H. Hough\altaffilmark{6}, and J. -I. Morino\altaffilmark{7}}

\altaffiltext{1}{Institute of Astrophysics and Planetary Sciences, 
Ibaraki University, Bunkyo 2-1-1, Mito, Ibaraki 310-8512, Japan.}
\altaffiltext{2}{momose@mito.ipc.ibaraki.ac.jp}
\altaffiltext{3}{National Astronomical Observatory of Japan, Osawa, Mitaka, 
Tokyo 181-8588, Japan.}
\altaffiltext{4}{Mizusawa Astrogeodynamics Observatory, Mizusawa, 
Iwate 023-0861, Japan.}
\altaffiltext{5}{Joint Astronomy Centre, 660 North A`oh\=ok\=u Place, 
University Park, Hilo, HI 96720.}
\altaffiltext{6}{Department of Physical Sciences, University of 
Hertfordshire, Hatfield, Hertfordshire AL10 9AB, UK.}
\altaffiltext{7}{Nobeyama Radio Observatory, Nobeyama, Minamisaku, Nagano 
384-1305, Japan.}

\begin{abstract}

Imaging polarimetry of the 850 $\mu$m continuum emission 
in the NGC 7538 region, obtained with the SCUBA Polarimeter, is 
presented. The polarization map is interpreted in terms of 
thermal radiation by magnetically aligned dust grains. Two 
prominent cores associated with IRS 1 and IRS 11, IRS 1(SMM) and IRS 11(SMM), 
are found in the surface brightness map. Although these cores look 
similar in surface brightness, their polarization shows striking 
differences. In IRS 11(SMM), the polarization vectors are extremely well 
ordered, and the degrees of polarization are quite high with an average 
of $\sim$3.9 \%. In IRS 1(SMM), on the other hand, the directions of 
polarization vectors are locally disturbed, and the degrees of 
polarization are much lower than those of IRS 11(SMM). These differences 
suggest that small scale fluctuations of the magnetic field are more 
prominent in IRS 1(SMM). This can be interpreted in terms of the difference 
in evolutionary stage of the cores. Inside IRS 1(SMM), which seems 
to be at a later evolutionary stage than IRS 11(SMM), substructures such 
as subclumps or a cluster of infrared sources have already 
formed. Small scale fluctuations in the magnetic field could have 
developed during the formation of these substructures. 
The distribution of magnetic field directions derived from our 
polarization map agrees well with those of molecular outflows 
associated with IRS 1(SMM) and IRS 11(SMM). Comparisons of energy 
densities between 
the magnetic field and the outflows show that the magnetic field 
probably plays an important role in guiding the directions of the outflows. 

\end{abstract}


\section{Introduction}

Outflows and jets are ubiquitous in young stellar objects (YSOs). 
Since the discovery of the first bipolar outflows in the 1980s 
(Snell et al. 1980), their study has become one of the central topics 
of star formation. It is suspected that outflows play a crucial 
role in removing angular momentum and magnetic fields and so 
are natural consequences of star formation processes (Shu, Adams, 
\& Lizano 1987). However, in spite of the major developments in 
revealing the dynamics of a large number of outflows by 
radio/millimeter-wave observations, the details of driving and collimating 
mechanisms of bipolar outflows are still controversial even if the 
magnetohydrodynamical models are in consideration (see a review by 
K\"onigl \& Ruden 1993). One of the important reasons why we have not 
been able to confidently judge these models with observations 
is the lack of information of the magnetic field structure near YSOs. 
The regions near YSOs are usually highly obscured by the embedding 
molecular cloud as well as the circumstellar matter such as envelopes 
or cloud cores. Therefore, conventional ``interstellar polarimetry'' 
at optical or near-infrared wavelengths cannot be employed to determine 
the field near YSOs. The most direct technique for mapping the magnetic 
fields near YSOs is to observe the polarized emission from dust at far-infrared 
or submillimeter wavelengths. Such techniques have become very efficient and 
powerful with use of large telescopes and array detectors that enable 
imaging polarimetry rather than point-by-point polarization mapping
(e.g., Hildebrand 1996; Davis et al. 2000; Matthews \& Wilson 2000; 
Schleuning et al. 2000). 

This paper presents polarization maps of the 850 $\mu$m 
continuum emission in the NGC 7538 region with the SCUBA Polarimeter 
(SCUPOL). The NGC 7538 molecular cloud is an active site of high-mass 
star formation containing several point-like infrared sources 
(Werner et al. 1979), and there are three prominent molecular outflows 
associated with these infrared sources (Kameya et al. 1989). Two remarkable
submillimeter/infrared sources, IRS 1 and IRS 11, are located in our mapped 
area. The inclination angle of the gaseous ring 0.13 pc in radius 
around IRS 1 has been estimated to be $37 ^{\circ}$ from high-resolution 
imaging (Kawabe et al. 1992), hence the outflow associated with IRS 1 is 
likely $\sim 30 ^{\circ}$ from the line of sight. The inclination angle of 
the IRS 11 outflow is less certain, but its spatial distribution is similar 
to that of the IRS 1 outflow (Kameya et al. 1989, 1990a), 
suggesting it is also around $30^{\circ}$. 
Point-by-point polarization measurements of the submillimeter 
continuum emissions from IRS 1 and IRS 11 were made by Flett \& Murray 
(1991) and Minchin \& Murray (1994). Their measurements suggested that 
the magnetic fields near these infrared sources are parallel to 
the directions of their outflows. Our polarization maps obtained with 
SCUPOL have revealed the internal magnetic field structure of the cloud 
cores surrounding IRS 1 and IRS 11 with unprecedented details, allowing 
us to discuss the relationship between the magnetic field and the 
outflows associated with infrared sources in this region. 

The outline of this paper is as follows. Our 
observations and the procedure of data reduction are described 
in $\S$2. In $\S3$ we present our polarization/magnetic field maps, 
and compare them with outflow distribution. The physical connection 
of the magnetic field structure with the evolution of cores and the 
outflows is discussed in $\S4$. 

\section{Observations and Data Reduction}

\subsection{Observations and Data}

The data were obtained with the James Clerk Maxwell Telescope 
(JCMT) atop Mauna Kea in Hawaii on 1998 July 18 (HST). The beam 
size was 14$''$ (FWHM) at 850 $\mu$m. Polarimetry was conducted 
with a rotating halfwave plate and a fixed wiregrid analyzer in 
front of the SCUBA (Submillimetre Common-User Bolometer Array) window. 
Further details of the polarimeter, SCUPOL, are described by Greaves 
et al. (2000). Jiggle maps were obtained at waveplate positions 
22.5$^{\circ}$ apart, repeated 8 times so that a sufficient signal 
to noise ratio was achieved. The atmospheric optical depth at 225 GHz,  
as measured by a tipping radiometer at CSO, was about 0.17 and stable 
during the observations. The airmass was $\sim$1.4 and the seeing was 
0.3$''$ at 350 GHz. Chopping was 120$''$  EW.
Unfortunately, the bright and compact submillimeter source IRS 9, 
whose peak intensity is $\sim 25$ \% of the 
IRS 1 peak, was located in the reference field, and some 
parts of the resultant maps of surface brightness and polarization 
vectors were corrupted. We shall describe its effects in detail when 
we present our results in $\S$ 3. 

\subsection{Reduction}

The ORAC-DR reduction program package was used for flat-fielding, 
extinction correction, sky-noise removal, bad-pixel removal, and rebinning. 
The sky removal was made from the median of the signal from the outermost 
bolometers that seemed emission-free in a short integration map. 
The polarization calibration including subtraction of 
instrumental polarization of $\sim$1 \% and debiasing, as well 
as plotting polarization vectors were also made using this package. 
The pixel size of the surface brightness map and the maps
of the Stokes $Q$ and $U$ parameters were originally taken to be 3$''$, 
but the polarization vectors were derived from data binned into $9''$ 
pixels. The resultant polarization data for positions 
at which the flux density is positive and the polarization degree 
is less than 10 \% are given in Table 1.
Flux calibration was made using the brightness at the position of IRS 1, 
19Jy per 17.5$''$ beam (Sandell 1994), corresponding to 493 mJy per 
$3''$ pixel in the surface brightness map.
Sky noise removal using the outermost bolometers causes a systematic 
error in flux measurement, but its amount is estimated to be only $\sim$
15 mJy per $3''$ pixel.  

\section{Results}

\subsection{850 $\mu$m Continuum Emission}

Figure 1 shows the surface brightness map of the 850 $\mu$m continuum 
emission. Strong submillimeter continuum emissions associated with 
IRS 1 and IRS 11 are clearly seen. Each core is resolved with the 
SCUBA beam and their shape is non-spherical, but being
elongated with several protrusions. The northern core (hereafter denoted by 
IRS 1(SMM)), which contains IRS 1, is associated with three protrusions to the 
east, to the west and to the southwest. These positionally coincide 
with the local peaks (CS2, CS4 and CS5, respectively) in a high resolution 
(6$''$ beam) map of the CS ($J=2-1$) line emission by Kawabe et al. (1992). 
The southern core (hereafter denoted by IRS 11(SMM)), which contains IRS 11 and 
has been much less studied, is associated with three protrusions to the east, 
to the southwest and to the northwest. 

Previous observations with submillimeter dust continuum as well as 
with the HCN and HCO$^{+}$ lines (Minchin \& Murray 1994; Cao et al. 1993; 
Kameya et al. 1990a and unpublished data) revealed faint emissions connecting 
two cores associated with IRS 1 and IRS 11. Although the existence of such 
emissions is also suggested in our surface brightness map, the position 
corresponding to IRS 9 in the reference beam is located just between the 
two cores (see Figure 1), corrupting part of the emissions connecting 
these two cores. Previous submillimeter continuum observations with the JCMT 
have revealed that the emissions associated with IRS 9 is so compact that  
the corrupted regions are expected to be approximately one beam 
(Greaves et al., unpublished data). Therefore, the spatial distributions of 
the core emissions in Figure 1 are hardly distorted, while detailed structure 
of the emissions between the cores cannot be discussed in this paper.

The total flux density of the continuum emission is 111 Jy, and the flux 
densities for IRS 1(SMM) and IRS 11(SMM) measured in the regions where 
the brightness is greater than 58.5 mJy per $3''$ pixel are 65 Jy and 29 Jy 
respectively. Assuming that the continuum emission is 
optically thin and the dust temperature $T_{d}$ is uniform, one can estimate 
the cloud mass by 
\begin{equation}
M = \frac{F_{\lambda}d^{2}}{\kappa_{\lambda}B_{\lambda}(T_{d})}
\end{equation}
where $\kappa_{\lambda}$ is the mass absorption coefficient by dust grains, 
$B_{\lambda}(T)$ is the Planck function, $F_{\lambda}$ is the flux density 
of the continuum emission, and $d$ is the distance to these cores. With 
$T_{d} = 25$ K (see Dickel, Dickel, \& Wilson 1981), $d=2.7$ kpc 
(\textit{IAU Trans.}, \textbf{12B}, 351 [1964]) and 
$\kappa_{850\mu \scriptsize{\mbox{m}}} = 8.65 \times 10^{-3}$ cm$^{2}$ g$^{-1}$, 
which is obtained by the formula of Hildebrand (1983) with $\beta = 2$, 
the total mass is derived to be $6.7 \times 10^{3} M_{\odot}$, and for 
IRS 1(SMM) and IRS 11(SMM) to be $3.9 \times 10^{3} M_{\odot}$ and $1.8 
\times 10^{3} 
M_{\odot}$, respectively. The total mass described above is in 
fairly good agreement with that derived from the CS and C$^{34}$S ($J=1-0$) 
mapping observations by Kameya et al. (1986), though our measurement
missed part of the faint emission components connecting the two cores. 

\subsection{850 $\mu$m Continuum Polarization and Magnetic Field}

Figures $2a$ shows the map of the 850 $\mu$m polarization vectors 
superposed on the surface brightness map. Since the 850 $\mu$m continuum 
emission from this region is thermal radiation from dust, the polarization 
is due to the emission from magnetically aligned dust grains (Davis 
\& Greenstein 1951). In Figure $2b$, the polarization vectors are rotated 
by 90$^{\circ}$ so that it directly shows the magnetic field direction 
projected on the sky. The polarization vectors in the main bodies of 
the cores systematically change their amplitudes and directions, indicating 
that the internal field structure of these two cores is successfully 
revealed. However, the contamination from IRS 9 in the reference field 
does not allow us to discuss the field structure in the regions 
between the cores. In addition, polarized emission components from 
side lobes of the beam, which cause error in polarization measurement
for fainter regions of extended sources, cannot be removed. 
Polarized components from side lobes can be as strong as 0.1 \% of the 
peak intensity within or near the field of view (Greaves 2000, SCUBA 
Polarimeter Commissioning Results in Semester 99B and 00A). We shall 
therefore use polarization data where the flux density is higher than 
10 \% of the IRS 1(SMM) peak when we give quantitative description of 
the field structure. 

The surface brightness map of the continuum emission gives an 
impression that the region is comprised of two similar cores, but the 
magnetic field map reveals clear contrast between IRS 1(SMM) and 
IRS 11(SMM). Therefore, we shall describe the details of the magnetic 
field structure of each region and its relationship with the outflow from 
the central sources.

\subsubsection{IRS 11(SMM)}

As seen in Figure $2b$, the magnetic field structure over IRS 11(SMM) 
is extremely well ordered. Figure $3a$ shows the histogram of the magnetic field 
direction in IRS 11(SMM) where the flux density is higher than 
10 \% of the IRS 1(SMM) peak (24 points). Prominent is the 170$^{\circ}$ 
component, which dominates the region to the south of IRS 11. As one moves to 
the north, its direction smoothly changes over IRS 11 and is pointed to the west. 
Simple histogram of the field direction does not reveal the coherent and 
gradual change of the field direction. Therefore we have calculated an 
average by rows and subtracted the average from each point of the rows. The 
resultant histogram of the local ``deviation'' of the field is shown in 
Figure $4a$. The histogram is symmetric around the peak with small variance, 
showing the coherence or uniformity of the magnetic field in IRS 11(SMM).

The smoothly bent but locally uniform magnetic field is in extremely good 
agreement with the morphology of the CO $J=3-2$ bipolar outflow (Kameya et 
al. 1990a), as shown in Figure $2b$. The CO outflow appears roughly bipolar 
to the south of IRS 11, consistent with the north-south field direction in 
this area. In particular, the outflow to the north, or the blueshifted 
component, gradually bends to the west, showing good correlation with the 
smoothly bent magnetic field. These facts suggest either that the magnetic 
field controls the directions of the outflows, or that the magnetic field 
structure is influenced by the outflow dynamics. We shall discuss further 
on this point in $\S$ 4.2. 

Figure $5a$ shows the histogram of the degrees of polarization. The degrees of 
polarization over IRS 11(SMM) are very high, ranging from 1 to 9\% with an 
average of 3.9\%. This polarization level is higher than the typical 
percentage of ``on-peak'' submillimeter polarization of star-forming 
regions observed with the same beam of the JCMT ($\sim 14''$, 
Tamura et al. 1999; Vall\'ee \& Bastien 2000) and is much higher than 
that of IRS 1(SMM) (see $\S$ 3.2.2). The degree and position angle 
of the polarized emission at the peak of IRS 11(SMM) obtained from 
our data are (3.0 $\pm$ 0.1)\% and $\sim$62$^{\circ}$. These values are in 
good agreement with the 800 $\mu$m polarization obtained by Flett \& Murray 
(1991) (3.8 $\pm$ 1.0 \% and 52 $\pm$ 8$^{\circ}$) or Minchin \& Murray (1994) 
(2.5 $\pm$ 0.2 \% and 58 $\pm$ 2$^{\circ}$) with the JCMT.

\subsubsection{IRS 1(SMM)}

The dominant magnetic field component over IRS 1(SMM) is the field running 
from northwest to southeast. The median of the distribution of the magnetic 
field direction in IRS 1(SMM) where the flux density is higher than 
10 \% of the IRS 1(SMM) peak (Figure $3b$; 40 points) is $\sim 150^{\circ}$. 
The degrees of polarization in this region range from 0 to 4\% with an average 
of 1.6 \% (Figure $5b$). This level of polarization has typically been 
observed in a number of star-forming regions (e.g., Vall\'ee \& Bastien 2000). 

It is noteworthy that the dominant magnetic field is consistent with the 
direction of the large-scale ($\sim$0.3 pc) CO bipolar outflow (Figure $2b$, see 
also Fischer et al. 1985; Scoville et al. 1986; Kameya et al. 1989). The 
blueshifted component is to the northwest of IRS 1 and the redshifted component 
is to the southeast, both in the position angle of $\sim 135^{\circ}$. 
In contrast, the radio continuum emission originating from jets shows a 
gradual change of its direction: from $\sim$180$^{\circ}$ on 0.005 pc 
scale to $\sim$165$^{\circ}$ on 0.03 pc scale (Campbell 1984). 

In addition to the dominant magnetic field, there are several fine structures 
seen in Figures $2a$ and $2b$. Most interesting is the presence of two 
small polarization regions on the opposite sides of IRS 1: 25$''$ southwest of 
IRS 1 where the degree of polarization is as small as $0.2 \%$ and northeast 
of IRS 1 where the degrees of polarization are $\sim 0.5\%$. This reminds us 
of the ``polarization disks'' that are often observed in the near-IR 
polarization maps of star-forming regions and are indicative of the disk 
plane and therefore perpendicular to the outflow direction 
(e.g., Tamura et al. 1991; Lucas \& Roche 1998). 
The direction of the ``submillimeter polarization disk'' in IRS 1(SMM) roughly 
coincides with the major axis of the disklike structure observed with molecular 
emissions (e.g., Scoville et al. 1986; Kawabe et al. 1992), although the 
origin of such polarization pattern at submillimeter wavelengths is not 
clear (for the case of a filamentary cloud, see Fiege \& Pudritz 2000). 

Another intriguing feature of the magnetic field is the relative non-uniformity 
of the field over IRS 1(SMM), when compared with IRS 11(SMM). Although the 
total change of the magnetic field across IRS 11(SMM) is large (as large as 
$\sim 60^{\circ}$), the field direction change is gradual. In contrast, 
the overall field change in IRS 1(SMM) is not so large, but the field is 
locally more chaotic. This behavior is shown in the histogram of the 
local ``deviation'' of the field constructed in the same manner as that for 
IRS 11 (Figure $4b$). The histogram is asymmetric around the peak with 
larger variance.

The degree and position angle of the polarized emission at IRS 1 obtained from 
our data are (3.0 $\pm$ 0.2)\% and $\sim$ 87$^{\circ}$. 
Flett \& Murray (1991) observed  800 $\mu$m  polarization vector at the position 
$\sim 9''$ north of IRS 1. Their polarization vector, $0.7 \pm 0.3$\% and 
$52 \pm 15^{\circ}$, is not consistent with our data ($2.1 \pm 0.3$ \% and  $72 
\pm 3^{\circ}$, see no. 108 in Table 1), but it agrees with our data for 
the position 9$''$ east of their observed position 
($0.8 \pm 0.3$ \% and  $52 \pm 10^{\circ}$, see no. 107 
in Table 1). The possible pointing error in their measurements as well as 
their larger beamsize (19$''$ (FWHM), see Minchin \& Murray (1994)) might cause 
the above inconsistency. 

\section{Discussion}

\subsection{Relationship between Core Evolution and Magnetic Field Structure}

Previous observations of the NGC 7538 region revealed that IRS 1(SMM) and 
IRS 11(SMM) commonly show features that can be regarded as the signs of ongoing 
star formation: huge far-infrared luminosities indicating the existence 
of embedded energy sources (Werner et al. 1979), bipolar outflows probably 
powered by those embedded young stellar objects (Kameya et al. 1989), 
and H$_{2}$O maser spots closely associated with ultracompact H 
{\small II} regions 
(Kameya et al. 1990b). Infrared images taken at wavelengths shorter than 20 
$\mu$m, however, manifest remarkable differences between these two cores. The 
point sources in IRS 1(SMM), as well as their surrounding nebulosity, can easily 
be identified at 2.2 $\mu$m, but only a faint extended emission is found in 
IRS 11(SMM) at the same wavelength (e.g., Davis et al. 1998). Even at $10 - 20 
\mu$m there is no detectable source in IRS 11(SMM) (Werner et al. 1979). These 
facts suggest that young stellar objects in IRS 11(SMM) are more deeply 
embedded compared with those in IRS 1(SMM), and that IRS 11(SMM) is in an earlier 
evolutionary stage of star formation. It is therefore expected that the 
comparison of polarization maps between these two cores will provide 
information about the change of internal magnetic-field structure during 
the course of core evolution. 

The path of core evolution depends on whether its initial state is 
magnetically subcritical or supercritical. Although it is difficult to 
precisely measure the mass-to-flux ratio in cores with observations, the 
most massive cores seem to be in magnetically supercritical state, in which 
the magnetic force is not strong enough to support the core (Bertoldi \& 
McKee 1992; see also Crutcher 1999). Scott \& Black (1980) made detailed 
numerical calculations of the collapse of a magnetically supercritical core 
and found that the collapse occurs preferentially along the field direction 
to form a disklike subregion with higher density. The disklike subregion is 
likely to fragment further into smaller subclumps because this region has the 
same ratio of surface-density to magnetic-flux as the entire cloud and is 
still in a magnetically supercritical state (Scott \& Black 1980; see 
also Mestel 1985; Shu et al. 1987). Since the interstellar medium 
couples well with magnetic fields (unless the hydrogen number density is 
greater than 10$^{11}$ cm$^{-3}$ (Nakano \& Umebayashi 1986)), the magnetic 
field structure inside a core must strongly be affected by dynamical 
processes in the core described above. It is expected that small scale 
fluctuations of the magnetic field grow as substructures inside the core 
form.

Our polarization map shows striking differences between IRS 1(SMM) and 
IRS 11(SMM). The directions of polarization vectors in IRS 1(SMM) are locally 
disturbed while those in IRS 11(SMM) are well ordered, suggesting that small 
scale fluctuations of magnetic field are more prominent in IRS 1(SMM). The 
non-uniformity of polarization degree in IRS 1(SMM), which is not found in 
IRS 11(SMM), can also be explained by small scale fluctuations of the magnetic 
field; the observed polarization degree gets smaller if the field direction 
significantly changes within the beamsize ($\sim$15$''$ = 0.2 pc). Such 
differences in magnetic field structure can be interpreted in terms of the 
difference in evolutionary stage between these cores. Previous observations 
have shown that the formation of substructures inside IRS 1(SMM) is highly 
progressed: there is disklike structure with a radius of around 0.3 pc 
(Scoville et al. 1986), and inside this structure several self-gravitating 
subclumps with a few tens of solar masses and  0.13 pc ($\sim 10''$) in size 
have been found by aperture 
synthesis observations (Kawabe et al. 1992). Moreover, a cluster of 
infrared sources has already formed within the central regions of IRS 1(SMM) 
(Wynn-Williams, Becklin, \& Neugebauer 1974a). 
The formation of these substructures in 
IRS 1(SMM) would cause the development of small scale fluctuations in the 
magnetic field. Small scale variations of magnetic field direction have also 
been found in the Orion BN/KL region, in which disklike structures and 
subclumps have also already formed (Burton et al. 1991; Chrysostomou et al. 
1994). On the other hand, IRS 11(SMM) is expected to be in an earlier 
evolutionary stage compared with IRS 1(SMM). It is therefore likely that 
fragmentation in IRS 11(SMM) has not yet highly developed, and that the 
magnetic field structure remains well ordered. 
Recent submillimeter observations of 
prestellar cores in the OMC-3 region, which is in an earlier stage of 
evolution compared with the Orion BN/KL region (Chini et al. 1997),
have revealed well ordered and high-degree of polarization vectors 
similar to those in the IRS 11(SMM) of NGC 7538 (Matthews \& Wilson 2000). 
We should keep in mind, however, that there are no observations which show 
the lack of prominent substructure inside IRS 11(SMM). Higher resolution 
observations of molecular gas toward IRS 11(SMM) are required in order to 
clearly reveal the relationship between the physical evolution of cores 
and magnetic field structure. 

\subsection{Comparisons between the Polarization Map and Molecular Outflows}

The magnetic field directions derived from our polarization map are 
consistent with those of molecular outflows associated with IRS 1(SMM) and 
IRS 11(SMM) (see Figure $2b$). There are two possible explanations for this 
result as follows: (i) the magnetic field strength in these cores is so 
strong that it controls the directions of the outflows, or (ii) the 
magnetic field strength is so weak that the field structure is influenced by 
the outflow dynamics. Comparisons of the energy density between the 
magnetic field and outflows are useful to judge which case is more 
plausible. If the magnetic field in each core has a greater energy density 
than the outflow, the former case is more plausible. Otherwise, the latter 
case is more plausible. Although this consideration might be too simplified, 
a more elaborate model for a \textit{low-mass} outflow by Hurka, 
Schmid-Burgk, \& Hardee (1999), who 
made numerical calculations of the deflection of high-velocity jets by 
ambient magnetic field, gave a similar criterion for the maximum velocity 
of a jet that can be deflected by the ambient magnetic field. 

The critical magnetic field strength at which the field has the same energy 
density as the outflows, 
$B_{\scriptsize{\mbox{flow}}}$, can be expressed by
\begin{equation}
B_{\scriptsize{\mbox{flow}}} = 
\left(\frac{8\pi E_{{\scriptsize{\mbox{flow}}}}}{V_{{\scriptsize{\mbox{flow}}}}}
\right)^{1/2}, 
\label{eqn:bflow}
\end{equation}
where $E_{{\scriptsize{\mbox{flow}}}}$ and $V_{{\scriptsize{\mbox{flow}}}}$ 
are the total kinetic energy and volume of the outflow, which can be estimated 
from outflow observations (e.g., Kameya et al. 1989). 
$B_{\scriptsize{\mbox{flow}}}$ for each core, as well as the physical parameters 
of the outflows, are summarized in Table 2. 
The magnetic field strengths in IRS 1(SMM) and IRS 11(SMM), on the other hand, 
can roughly be estimated from their column densities, though 
it cannot be derived directly from our observations. Using the data set 
compiled by Cructer (1999), who examined the relationships between the 
magnetic field strengths in molecular clouds and their 
physical parameters, one can find the following relationship between the field 
strengths and column densities with its correlation coefficient of 0.85:
\begin{equation}
B  = \frac{\pi}{2} B _{\scriptsize{\mbox{los}}}
= 229 \left(\frac{N(\mbox{H}_{2})}{10^{23}\mbox{cm}^{-2}}\right)^{0.95} 
(\mu\mbox{G}),
\label{eqn:crutcher}
\end{equation}
where $B _{\scriptsize{\mbox{los}}}$
is magnetic field strength obtained by Zeeman measurements and 
$N(\mbox{H}_{2})$ is column density of a cloud. 
Using this equation the magnetic field strengths in IRS 1(SMM) and IRS 11(SMM) 
($N(\mbox{H}_{2}) \approx 3.5 \times 10^{23}$ cm$^{-2}$, see row (4) in Table 2) 
are estimated to be $\sim$ 750 $\mu$G, which is greater than 
$B_{\scriptsize{\mbox{flow}}}$ (see row (8) in Table 2). 
It is therefore plausible that the magnetic field guides the 
outflows on a larger scale. The field strength 
estimated from equation (\ref{eqn:crutcher}) is smaller than the critical field 
strength of each core at which the magnetic 
force is comparable to the gravitational force (see row (9) in Table 2). 
This is consistent with the fact that young stars have already formed in 
these cores, though the mechanical support from turbulent pressure is not 
taken into account when the critical field strength in row (9) is estimated. 

In addition to the large-scale outflows, the magnetic field direction in 
IRS 1(SMM) seems to be correlated with a jet on a smaller scale: 
although the dominant magnetic field in IRS 1(SMM) is running from northwest 
to southeast, the field near the emission peak is almost in the north-south 
direction (see Figure $2b$) which is consistent with the direction of the 
0.03 pc-scale jet ejected from IRS 1 (Campbell 1984). This fact could suggest 
that the magnetic field plays an important role in collimating small-scale 
jets, as has been discussed in many theoretical studies (e.g., K\"onigl 
\& Ruden 1993). However, the angular resolution of 
our map is not sufficient to discuss this topic in detail. Polarimetry with 
higher angular resolution is desired to obtain more convincing evidence for 
the small scale collimation. 

One may suspect that the small scale fluctuations of the 
polarization vectors in IRS 1(SMM) could be related to the outflows ejected 
from young stellar objects inside the core. However, this seems unlikely 
because the physical parameters of the outflow in IRS 1(SMM) are not so 
different from those of the outflow in IRS 11(SMM) (see Table 2) where 
the small scale fluctuations of magnetic field are not found. 

\subsection{Relationship between Cores and Galactic Magnetic Field}

The direction of the dominant magnetic field component in IRS 1(SMM) and 
IRS 11(SMM) (position angle $\sim 160^{\circ}$; see Figures $3a$ \& $3b$) 
does not align with the interstellar 
polarization which can be regarded as the direction of galactic magnetic 
field (position angle $\sim 60^{\circ}$, see Dyck \& Lonsdale (1979)). 
This fact may suggest that non-magnetic forces such as self-gravity of the
interstellar medium or compression by the nearby H {\small II} region (Kameya \& 
Takakubo 1988) regulate the formation processes of these cores, and that 
the magnetic field inside the cores has experienced complete restructuring 
during their formation. 

\section{Conclusions}

We have made imaging polarimetry of the 850 $\mu$m continuum emission 
in the NGC 7538 region with the SCUBA Polarimeter mounted on the JCMT. 
The polarization maps are interpreted in terms of thermal 
radiation by magnetically aligned dust grains. Our conclusions are 
summarized as follows: 

1. Two prominent cores associated with IRS 1 and IRS 11 (IRS 1(SMM) and 
IRS 11(SMM)) are found in the 
surface brightness map of the continuum emission. The total 
cloud mass derived from our surface brightness map is $6.7 \times 10^{3} 
M_{\odot}$. 

2. The polarization map shows striking difference between IRS 1(SMM) and 
IRS 11(SMM). In IRS 11(SMM), the polarization vectors are extremely well ordered, 
and the degrees of polarization are very high, ranging from 1 to 9 \% 
with an average of 3.9 \%. In IRS 1(SMM), on the other hand, the directions 
of polarization vectors are locally disturbed, and the degrees of 
polarization range from 0 to 4 \% with an average of 1.6 \%, which is 
much lower than that of IRS 11(SMM). These differences suggest that small 
scale fluctuations of magnetic field are more prominent in IRS 1(SMM). 

3. Such differences in magnetic field structure can be interpreted 
in terms of a difference in evolutionary stage between these cores.
Inside IRS 1(SMM) substructures such as subclumps or a cluster 
of infrared sources have already formed. Small scale 
fluctuations in the magnetic field can develop during the 
formation of these substructures. On the other hand, IRS 11(SMM) has no 
detectable source at wavelengths shorter than 20 $\mu$m and seems 
to be in an earlier evolutionary stage compared with IRS 1(SMM). It is
therefore likely that fragmentation in IRS 11(SMM) has not yet 
sufficiently progressed, and that the magnetic field structure remains 
well ordered. 

4. The magnetic field directions derived from our polarization map 
agree well with those of molecular outflows associated with IRS 1(SMM) 
and IRS 11(SMM). Although this fact suggests either that the magnetic 
field controls the directions of the outflows, or that the magnetic 
field structure is influenced by the outflow dynamics, comparisons of 
energy densities between the magnetic field and the outflows show that 
the former case is more plausible. 

\acknowledgments

The JCMT is operated by the Joint Astronomy Centre,
on behalf of the UK Particle Physics and Astronomy
Research Council, the Netherlands Organization for
Pure Research, and the National Research Council of
Canada. We would like to thank N. Kaifu, K. Kodaira,
S. Sato, and A. Boksenberg for promoting the UK-Japan
collaboration. This work was supported in part by the Grant in-aid
for Scientific Research of Ministry of Education, Science,
Sports and Culture, and by the Japan Society of the Promotion
of Science.

\clearpage

\figcaption[figure1.eps]{The surface brightness map of the 850 $\mu$m continuum 
emission in the NGC 7538 region. The pixel size is $3'' \times 3''$ and the 
contour spacing is 19.5 mJy pixel$^{-1}$, starting at 9.75 mJy beam$^{-1}$. The 
dashed lines indicate the level of $- 9.75$ mJy pixel$^{-1}$. The heavy line
indicates the field of view. Triangles and squares show the positions of infrared 
sources (Werner et al. 1979) and OH masers (Wynn-Williams, Werner, 
\& Wilson 1974) respectively. The grey circle with a black cross in the map 
indicates the one-beam (FWHM) area around IRS 9 in the reference beam. 
\label{fig1}}

\figcaption[figure2.eps]{($a$) Polarization $E$ vectors of the 
850 $\mu$m continuum emissions (thick lines) superposed on the 
surface brightness map shown in 
Figure 1 (grey scale). In total, 122 data points are plotted. 
The length of each line is proportional to the polarization degree. 
($b$) Magnetic field directions derived from the polarization vectors in ($a$) 
(black and violet lines) superposed on the distribution of CO($J=3-2$) 
high-velocity outflows obtained by Kameya et al. (1990a) (contours) and 
the 850 $\mu$m surface brightness map (grey scale). The black lines indicate 
field directions inside the regions where the flux density is higher than 
10 \% of the IRS 1(SMM) peak while the violet lines indicate those outside 
the regions. Only the data shown by black lines are used when the
quantitative description of the field structure is made.
The blue and red contours show the distribution of the CO($J=3-2$) 
emissions integrated over $v_{lsr} = -74$ to $-64$ km s$^{-1}$ and 
$v_{lsr} = -54$ to $-44$ km s$^{-1}$, respectively. The grey circle with a 
black cross in each map indicates the one-beam (FWHM) area around IRS 9 in 
the reference beam. \label{fig2}}

\figcaption[figure3.eps]{($a$) Histogram of the magnetic field directions
 (i.e., the position angles of 850 $\mu$m polarization plus 90$^{\circ}$) 
 in IRS 11(SMM). ($b$) Same as ($a$) but in IRS 1(SMM).}

\figcaption[figure4.eps]{($a$) Histogram of the deviations of the magnetic 
field direction from an average of a local magnetic field in IRS 11(SMM). 
($b$) Same as ($a$) but in IRS 1(SMM).}

\figcaption[figure5.eps]{($a$) Histogram of the polarization degrees 
in IRS 11(SMM). ($b$) Same as ($a$) but in IRS 1(SMM).}

\clearpage

\begin{deluxetable}{crrrrrrr}
\tabletypesize{\scriptsize}
\tablecaption{Flux densities, polarization degrees and its position angles \label{tbl-1}}
\tablewidth{0pt}
\tablehead{
\colhead{Position} & \colhead{$\Delta$R.A.\tablenotemark{a}}   & 
\colhead{$\Delta$Dec. \tablenotemark{a}}   &
\colhead{$F_{\nu}$\tablenotemark{b}}     & \colhead{$P$}  &
\colhead{$\Delta P$}   & \colhead{P.A.}
 & \colhead{$\Delta$P.A.} \\
\colhead{No.} & \colhead{$('')$}   & 
\colhead{$('')$}   &
\colhead{(mJy) }     & \colhead{(\%)}  &
\colhead{(\%)}   & \colhead{($^{\circ}$)}
 & \colhead{($^{\circ}$)}
}
\startdata
1 & -16.5 & -73.5 & 156 & 4.32 & 1.12 & -26.6 & 6.5 \\
2 & -25.5 & -73.5 & 267 & 5.18 & 0.85 & -71.7 & 4.5 \\
3 & -34.5 & -73.5 & 292 & 3.68 & 0.58 & -25.4 & 4.7 \\
4 & -43.5 & -73.5 & 284 & 3.58 & 0.81 & -11.2 & 6.3 \\
5 & 37.5 & -64.5 & 56 & 7.04 & 2.44 & 89.7 & 9.7 \\
6 & 28.5 & -64.5 & 162 & 7.58 & 0.81 & -87.1 & 3.0 \\
7 & 19.5 & -64.5 & 217 & 8.75 & 0.64 & 85.3 & 2.0 \\
8 & 10.5 & -64.5 & 292 & 6.99 & 0.55 & 89.2 & 2.1 \\
9 & 1.5 & -64.5 & 399 & 5.59 & 0.45 & -86.3 & 2.0 \\
10 & -7.5 & -64.5 & 447 & 6.24 & 0.42 & -82.6 & 1.8 \\
11 & -16.5 & -64.5 & 457 & 2.55 & 0.33 & -75.3 & 3.8 \\
12 & -25.5 & -64.5 & 458 & 1.10 & 0.27 & -72.6 & 6.5 \\
13 & -34.5 & -64.5 & 389 & 2.25 & 0.31 & 15.9 & 4.5 \\
14 & -43.5 & -64.5 & 239 & 6.73 & 0.56 & 21.1 & 3.0 \\
15 & 37.5 & -55.5 & 210 & 3.76 & 0.71 & -73.4 & 4.8 \\
16 & 28.5 & -55.5 & 365 & 6.26 & 0.58 & -84.7 & 2.6 \\
17 & 19.5 & -55.5 & 560 & 8.32 & 0.45 & 82.8 & 1.5 \\
18 & 10.5 & -55.5 & 843 & 5.16 & 0.29 & 85.5 & 1.6 \\
19 & 1.5 & -55.5 & 1439 & 1.80 & 0.22 & 72.9 & 3.7 \\
20 & -7.5 & -55.5 & 1557 & 2.57 & 0.23 & 85.6 & 2.5 \\
21 & -16.5 & -55.5 & 980 & 1.75 & 0.23 & 76.4 & 3.7 \\
22 & -25.5 & -55.5 & 614 & 2.59 & 0.26 & 75.6 & 2.8 \\
23 & -34.5 & -55.5 & 330 & 1.33 & 0.47 & -27.1 & 9.3 \\
24 & 46.5 & -46.5 & 95 & 7.59 & 1.51 & -30.5 & 5.1 \\
25 & 37.5 & -46.5 & 303 & 8.65 & 0.72 & -78.3 & 2.2 \\
26 & 19.5 & -46.5 & 883 & 7.74 & 0.54 & 76.4 & 1.9 \\
27 & 10.5 & -46.5 & 1727 & 4.90 & 0.21 & 77.1 & 1.2 \\
28 & 1.5 & -46.5 & 3200 & 3.00 & 0.09 & 62.4 & 0.9 \\
29 & -7.5 & -46.5 & 2917 & 2.63 & 0.10 & 61.5 & 1.1 \\
30 & -16.5 & -46.5 & 1414 & 1.79 & 0.26 & 61.0 & 4.1 \\
31 & -25.5 & -46.5 & 558 & 5.16 & 0.44 & 83.3 & 2.4 \\
32 & 46.5 & -37.5 & 97 & 3.22 & 1.40 & 89.4 & 14.0 \\
33 & 37.5 & -37.5 & 286 & 6.72 & 1.66 & -60.5 & 6.9 \\
34 & 19.5 & -37.5 & 688 & 2.92 & 0.78 & 74.1 & 7.5 \\
35 & 10.5 & -37.5 & 1428 & 3.48 & 0.25 & 57.6 & 2.0 \\
36 & 1.5 & -37.5 & 2151 & 4.02 & 0.14 & 54.6 & 1.2 \\
37 & -7.5 & -37.5 & 1726 & 3.97 & 0.18 & 45.7 & 1.6 \\
38 & -16.5 & -37.5 & 836 & 3.61 & 0.27 & 30.5 & 2.1 \\
39 & -25.5 & -37.5 & 369 & 4.19 & 0.41 & 23.6 & 2.8 \\
40 & 10.5 & -28.5 & 530 & 5.62 & 0.43 & 58.7 & 2.0 \\
41 & 1.5 & -28.5 & 654 & 3.50 & 0.35 & 49.6 & 2.6 \\
42 & -7.5 & -28.5 & 641 & 4.32 & 0.26 & 37.3 & 1.7 \\
43 & -16.5 & -28.5 & 570 & 5.54 & 0.21 & 20.4 & 1.0 \\
44 & -25.5 & -28.5 & 314 & 5.47 & 0.42 & 16.9 & 2.2 \\
45 & -7.5 & -19.5 & 389 & 7.17 & 0.36 & 36.7 & 1.4 \\
46 & -16.5 & -19.5 & 534 & 5.19 & 0.24 & 24.5 & 1.2 \\
47 & -25.5 & -19.5 & 414 & 3.50 & 0.31 & 22.9 & 2.7 \\
48 & -34.5 & -19.5 & 232 & 4.15 & 0.52 & 2.5 & 3.6 \\
49 & -43.5 & -19.5 & 127 & 7.88 & 0.97 & -0.7 & 4.0 \\
50 & -7.5 & -10.5 & 383 & 5.67 & 0.61 & 31.4 & 3.1 \\
51 & -16.5 & -10.5 & 527 & 2.12 & 0.31 & 53.6 & 4.2 \\
52 & -25.5 & -10.5 & 554 & 1.83 & 0.24 & 28.7 & 3.8 \\
53 & -34.5 & -10.5 & 452 & 1.16 & 0.30 & 32.6 & 7.2 \\
54 & -43.5 & -10.5 & 294 & 1.51 & 0.49 & 35.1 & 9.0 \\
55 & -52.5 & -10.5 & 131 & 7.07 & 0.94 & 6.4 & 4.1 \\
56 & 19.5 & -1.5 & 224 & 4.68 & 0.63 & 45.2 & 3.9 \\
57 & 10.5 & -1.5 & 402 & 4.76 & 0.34 & 41.8 & 2.1 \\
58 & 1.5 & -1.5 & 611 & 4.64 & 0.27 & 35.3 & 1.6 \\
59 & -7.5 & -1.5 & 778 & 3.12 & 0.21 & 41.3 & 1.6 \\
60 & -16.5 & -1.5 & 797 & 1.56 & 0.21 & 54.5 & 3.4 \\
61 & -25.5 & -1.5 & 663 & 2.51 & 0.21 & 8.8 & 2.5 \\
62 & -34.5 & -1.5 & 509 & 1.50 & 0.29 & -11.1 & 5.4 \\
63 & -43.5 & -1.5 & 282 & 0.48 & 0.50 & 34.5 & 20.7 \\
64 & 28.5 & 7.5 & 188 & 4.28 & 0.64 & 36.0 & 4.3 \\
65 & 19.5 & 7.5 & 428 & 1.73 & 0.30 & 45.9 & 5.5 \\
66 & 10.5 & 7.5 & 747 & 2.90 & 0.18 & 43.3 & 2.0 \\
67 & 1.5 & 7.5 & 1026 & 2.21 & 0.15 & 40.2 & 1.9 \\
68 & -7.5 & 7.5 & 1277 & 1.33 & 0.13 & 53.6 & 2.5 \\
69 & -16.5 & 7.5 & 1195 & 1.01 & 0.11 & 50.3 & 3.6 \\
70 & -25.5 & 7.5 & 800 & 1.20 & 0.19 & 22.8 & 4.2 \\
71 & -34.5 & 7.5 & 482 & 2.32 & 0.30 & -4.7 & 3.4 \\
72 & -43.5 & 7.5 & 245 & 1.90 & 0.56 & -1.2 & 7.3 \\
73 & 37.5 & 16.5 & 146 & 6.79 & 0.75 & 35.3 & 3.1 \\
74 & 28.5 & 16.5 & 372 & 2.00 & 0.32 & 61.0 & 4.8 \\
75 & 19.5 & 16.5 & 781 & 1.11 & 0.20 & 75.7 & 5.3 \\
76 & 10.5 & 16.5 & 1273 & 1.42 & 0.17 & 58.4 & 3.5 \\
77 & 1.5 & 16.5 & 1839 & 1.25 & 0.12 & 67.7 & 2.9 \\
78 & -7.5 & 16.5 & 1873 & 1.06 & 0.08 & 89.7 & 2.1 \\
79 & -16.5 & 16.5 & 1503 & 0.16 & 0.09 & -58.6 & 14.6 \\
80 & -25.5 & 16.5 & 901 & 0.72 & 0.17 & -15.8 & 6.5 \\
81 & -34.5 & 16.5 & 443 & 1.27 & 0.31 & 15.6 & 6.9 \\
82 & -43.5 & 16.5 & 217 & 4.39 & 0.63 & 1.8 & 4.0 \\
83 & 37.5 & 25.5 & 275 & 3.23 & 0.45 & 57.6 & 3.8 \\
84 & 28.5 & 25.5 & 596 & 2.10 & 0.24 & 56.4 & 3.5 \\
85 & 19.5 & 25.5 & 1185 & 1.67 & 0.18 & 46.4 & 3.3 \\
86 & 10.5 & 25.5 & 2287 & 1.39 & 0.17 & 66.8 & 3.5 \\
87 & 1.5 & 25.5 & 2967 & 1.60 & 0.13 & 85.9 & 2.5 \\
88 & -7.5 & 25.5 & 2405 & 1.22 & 0.11 & 79.3 & 2.6 \\
89 & -16.5 & 25.5 & 1712 & 0.38 & 0.10 & 88.1 & 6.9 \\
90 & -25.5 & 25.5 & 1064 & 1.48 & 0.16 & -23.9 & 3.0 \\
91 & -34.5 & 25.5 & 587 & 1.27 & 0.24 & -17.6 & 5.3 \\
92 & -43.5 & 25.5 & 227 & 3.98 & 0.56 & -22.1 & 4.0 \\
93 & 46.5 & 34.5 & 119 & 7.19 & 1.04 & 15.3 & 4.0 \\
94 & 37.5 & 34.5 & 418 & 2.02 & 0.29 & 31.7 & 4.1 \\
95 & 28.5 & 34.5 & 677 & 1.38 & 0.19 & 58.3 & 3.8 \\
96 & 19.5 & 34.5 & 1467 & 1.81 & 0.24 & 40.7 & 3.7 \\
97 & 10.5 & 34.5 & 3516 & 2.21 & 0.20 & 36.2 & 2.6 \\
98 & 1.5 & 34.5 & 4388 & 3.03 & 0.17 & 87.2 & 1.7 \\
99 & -7.5 & 34.5 & 3202 & 2.66 & 0.14 & 84.2 & 1.4 \\
100 & -16.5 & 34.5 & 1972 & 1.65 & 0.12 & 73.6 & 2.1 \\
101 & -25.5 & 34.5 & 1299 & 0.88 & 0.15 & 60.7 & 4.8 \\
102 & -34.5 & 34.5 & 696 & 0.41 & 0.22 & 2.1 & 13.6 \\
103 & -43.5 & 34.5 & 294 & 2.45 & 0.46 & -15.7 & 5.4 \\
104 & 37.5 & 43.5 & 309 & 4.52 & 0.51 & 32.4 & 3.2 \\
105 & 28.5 & 43.5 & 515 & 1.00 & 0.35 & 42.4 & 9.7 \\
106 & 19.5 & 43.5 & 1045 & 0.42 & 0.36 & 18.9 & 17.6 \\
107 & 10.5 & 43.5 & 2316 & 0.78 & 0.27 & 52.3 & 10.0 \\
108 & 1.5 & 43.5 & 2721 & 2.08 & 0.26 & 72.3 & 3.7 \\
109 & -7.5 & 43.5 & 1950 & 3.09 & 0.26 & 69.9 & 2.7 \\
110 & -16.5 & 43.5 & 1160 & 2.27 & 0.20 & 60.3 & 3.0 \\
111 & -25.5 & 43.5 & 828 & 1.73 & 0.25 & 54.8 & 4.3 \\
112 & -34.5 & 43.5 & 492 & 2.13 & 0.32 & 30.5 & 4.2 \\
113 & -43.5 & 43.5 & 276 & 4.95 & 0.54 & 2.3 & 2.6 \\
114 & 19.5 & 52.5 & 219 & 8.89 & 0.92 & -14.4 & 3.0 \\
115 & 10.5 & 52.5 & 561 & 3.00 & 0.58 & -36.4 & 5.6 \\
116 & 1.5 & 52.5 & 798 & 0.62 & 0.48 & 43.0 & 19.6 \\
117 & -7.5 & 52.5 & 630 & 2.39 & 0.50 & 58.5 & 6.4 \\
118 & -16.5 & 52.5 & 373 & 6.05 & 0.53 & 76.5 & 2.5 \\
119 & -25.5 & 52.5 & 228 & 4.95 & 0.67 & 76.8 & 3.9 \\
120 & -34.5 & 52.5 & 160 & 4.74 & 0.84 & 65.3 & 5.7 \\
121 & -43.5 & 52.5 & 116 & 4.40 & 1.22 & -42.6 & 8.1 \\
122 & -52.5 & 52.5 & 85 & 4.98 & 2.28 & 4.0 & 11.5 \\
\enddata

\tablenotetext{a}{Reference Position = (23$^{h}$11$^{m}$36.8$^{s}$, 
61$^{\circ}$11$'$15.3$''$) (B1950)}
\tablenotetext{b}{measured in $3\times 3$pixels ($9''\times 9''$)}
\end{deluxetable}

\clearpage

\begin{deluxetable}{lrr}
\tabletypesize{\normalsize}
\tablecaption{Physical Parameters and Critical Magnetic Strengths for 
IRS 1(SMM) and IRS 11(SMM) \label{tbl-2}}
\tablewidth{0pt}
\tablehead{
\colhead{ } & \colhead{IRS 1(SMM)}
 & \colhead{IRS 11(SMM)} \\
}
\startdata
(1) Spatial Extent (pc$^{2}$) $^{a}$ & 0.657 & 0.358 \\
(2) Mass ($M_{\odot}$) $^{a}$ & $3.9 \times 10^{3}$ & $1.8 \times 10^{3}$ \\
(3) Mean Column Density (g cm$^{-2}$) & 1.24 & 1.05 \\
(4) Mean Column Density ($N$(H$_{2}$) in cm$^{-2}$) & $3.7 \times 10^{23}$ & $3.2 
\times 10^{23}$ \\
(5) Outflow Energy (ergs) $^{b}$ & $6 \times 10^{46}$ & $4 \times 10^{46}$ \\
(6) Spatial Extent of Outflow (pc$^{2}$) $^{b}$ & 0.90 & 0.64 \\
(7) Expected Volume of Outflow (pc$^{3}$) $^{c}$ & 0.85 & 0.51 \\
(8) $B_{\scriptsize{\mbox{flow}}}$ ($\mu$G) $^{d}$ & 250 & 260 \\
(9) $B_{\scriptsize{\mbox{grav}}}$ ($\mu$G) $^{e}$ &
 $2.0 \times 10^{3}$ & $1.7 \times 10^{3}$ \\
\enddata

\tablenotetext{a}{derived from our observations. see $\S$ 3.1. }
\tablenotetext{b}{from Kameya et al. (1989).}
\tablenotetext{c}{estimated by (spatial extent)$^{1.5}$.}
\tablenotetext{d}{the critical field strength at which the field 
has the same energy density as the outflows, derived from equation 
(\ref{eqn:bflow}).}
\tablenotetext{e}{the critical field 
strength of the cloud at which the magnetic 
force is comparable to the gravitational force, derived from  
$B_{\scriptsize{\mbox{grav}}} = 2\pi \sqrt{G}
 \Sigma$, where $G$ is the gravitational constant and $\Sigma$ is 
 the column density of a cloud; see Nakano \& Nakamura (1978).}

\end{deluxetable}

\end{document}